\documentclass[10pt, conference, letterpaper]{IEEEtran}
\usepackage{cite}
\usepackage{graphicx}
\usepackage{amsfonts}
\usepackage{caption}
\usepackage{subcaption}
\usepackage{diagbox}
\usepackage{optidef}
\usepackage{amsmath}

\usepackage{tikz}
\usetikzlibrary{shapes}

\usepackage{stfloats}
\usepackage{multirow}
\usepackage{amsmath}
\usepackage{array}
\newcolumntype{C}[1]{>{\centering\let\newline\\\arraybackslash\hspace{0pt}}m{#1}}
\usepackage{url}
\usepackage{color,soul}
\usepackage{multirow}
\usepackage{hhline}
\usepackage{comment}
\usepackage{rotating}
\usepackage{optidef}
\usepackage{bbm}
\usepackage{enumitem}
\newcounter{descriptcount}

\usepackage{xcolor}
\usepackage{float}
\usepackage{caption}
\usepackage{subcaption}
\usepackage{stfloats}
\usepackage{url}
\usepackage{bbm}  
\usepackage{arydshln}
\usepackage{cite}
\usepackage{algorithm}
\usepackage{algpseudocode}
\usepackage[T1]{fontenc} 
\usepackage[utf8]{inputenc}
\usepackage{times}

\usepackage{amssymb}
\usepackage{amsmath}
\usepackage{lipsum}
\usepackage{enumitem}

\usepackage{bm}
\usepackage{algorithm}
\usepackage{algpseudocode}
\usepackage{braket}
\usepackage{threeparttable}

\usepackage[acronym]{glossaries} 
\newacronym{cwnd}{W}{Congestion Window}
\newacronym{aimd}{AIMD}{Additive Increase Multiplicative Decrease}
\newacronym{tcp}{TCP}{Transfer Control Protocol}
\newacronym{rtt}{RTT}{Round Trip Time}
\newacronym{cca}{CCA}{Congestion Control Algorithm}
\newacronym{bbr}{BBR}{Bottleneck Bandwidth and Round-trip propagation time}
\newacronym{fcc}{FCC}{Federal Communications Commission}
\newacronym{bdp}{BDP}{Bandwidth-Delay Product}
\newacronym{dtc}{DTC}{Direct to Cell}
\newacronym{mbb}{MBB}{Make Before Break}
\newacronym{3gpp}{3GPP}{3rd Generation Partnership Project}
\newacronym{5g}{5G}{Fifth Generation}
\newacronym{6g}{6G}{6th-Generation}
\newacronym{ai}{AI}{Artificial Intelligence}
\newacronym{cnn}{CNN}{Convolutional Neural Network}
\newacronym{cogsat}{CogSat}{Cognitive Satellite}
\newacronym{cr}{CR}{Cognitive Radio}
\newacronym{crsn}{CRSN}{Cognitive Radio Sensor Networks}
\newacronym{csa}{CSA}{Concurrent Spectrum Access}
\newacronym{csi}{CSI}{Channel State Information}
\newacronym{dlr}{DLR}{Deep Learning Reinforcement}
\newacronym{dsa}{DSA}{Dynamic Spectrum Allocation}
\newacronym{dsrc}{DSRC}{Dedicated Short-Range Communications}
\newacronym{fl}{FL}{Federated Learning}
\newacronym{geo}{GEO}{Geostationary Equatorial Orbit}
\newacronym{iobt}{IoBT}{Internet of Battle Things}
\newacronym{iot}{IoT}{Internet of Things}
\newacronym{isl}{ISL}{Inter-Satellite Link}
\newacronym{itu}{ITU}{International Telecommunication Union}
\newacronym{leo}{LEO}{Low Earth Orbit}
\newacronym{lstm}{LSTM}{Long Short-Term Memory}
\newacronym{madrl}{MADRL}{Multi-Agent Deep Reinforcement Learning}
\newacronym{mdp}{MDP}{Markov Decision Process}
\newacronym{meo}{MEO}{Medium Earth Orbit}
\newacronym{ml}{ML}{Machine Learning}
\newacronym{nfv}{NFV}{Network Function Virtualization}
\newacronym{ntn}{NTN}{Non-Terrestrial Networks}
\newacronym{osa}{OSA}{Opportunistic Spectrum Access}
\newacronym{pu}{PU}{Primary User}
\newacronym{qos}{QoS}{Quality of Service}
\newacronym{rl}{RL}{Reinforcement Learning}
\newacronym{rsma}{RSMA}{Rate Splitting Multiple Access}
\newacronym{satcom}{SatCom}{Satellite Communication}
\newacronym{siot}{SIoT}{Satellite Internet of Things}
\newacronym{sl}{SL}{Supervised Learning}
\newacronym{su}{SU}{Secondary User}
\newacronym{uav}{UAV}{Unmanned Aerial Vehicle}
\newacronym{usl}{USL}{Unsupervised Learning}
\newacronym{vsat}{VSAT}{Very Small Aperture Terminal}
\newacronym{wsn}{WSN}{Wireless Sensor Networks}
\newacronym{gps}{GPS}{Global Positioning Systems}
\newacronym{dsm}{DSM}{Dynamic Spectrum Management}
\newacronym{stn}{STN}{Satellite Terrestrial Network}
\newacronym{eirp}{EIRP}{Effective Isotropic Radiated Power}
\newacronym{ss}{SS}{Sectrum Sensing}
\newacronym{kpi}{KPI}{Key Performance Indicator}
\newacronym{ca}{CA}{Channel Availability}
\newacronym{sinr}{SINR}{Signal to Noise plus Interference Ratio}
\newacronym{snr}{SNR}{Signal to Noise Ratio}
\newacronym{inr}{INR}{Interference to Noise Ratio}
\newacronym{ipc}{IPC}{Interference Power Constraint}
\newacronym{drl}{DRL}{Deep Reinforcement Learning}
\newacronym{ddpg}{DDPG}{Deep Deterministic Policy Gradient}
\newacronym{rem}{REM}{Radio Environment Map}
\newacronym{sdn}{SDN}{Software Defined Network}
\newacronym{rnn}{RNN}{Recurrent Neural Network}
\newacronym{dl}{DL}{Deep Learning}
\newacronym{svm}{SVM}{Support Vector Machine}
\newacronym{dqn}{DQN}{Deep Q-Network}
\newacronym{dca}{DCA}{Dynamic Channel Allocation}
\newacronym{wan}{WAN}{Wide Area Network}
\newacronym{gnss}{GNSS}{Global Navigation Satellite System}
\newacronym{ppo}{PPO}{Proximal Policy Optimization}
\newacronym{vits}{VITS}{Very High Throughput Satellites}
\newacronym{llm}{LLM}{Large Language Model}
\newacronym{sdr}{SDR}{Software Defined Radio}
\newacronym{css}{CSS}{Cooperative Spectrum Sensing}
\newacronym{ieee}{IEEE}{Institute of Electrical and Electronics Engineers}
\newacronym{rssi}{RSSI}{Received Signal Strength Indicator}
\newacronym{wran}{WRAN}{Wireless Regional Area Network}
\newacronym{cstn}{CSTN}{Cognitive Satellite Terrestrial Network}
\newacronym{bs}{BS}{Base Stations}
\newacronym{noma}{NOMA}{Non-Orthogonal Multiple Access}
\newacronym{cicstn}{CI-CSTN}{Cooperative Integrated-Cognitive Satellite Terrestrial Network}
\newacronym{hts}{HTS}{High Throughput Satellite}
\newacronym{mimo}{MIMO}{Multiple Input Multiple Output}
\newacronym{fss}{FSS}{Fixed Satellite Service}
\newacronym{mss}{MSS}{Mobile Satellite Service}
\newacronym{bss}{BSS}{Broadcasting Satellite Service}
\newacronym{mifr}{MIFR}{Master International Frequency Register}
\newacronym{wrc}{WRC}{World Radiocommunication Conference}
\newacronym{ngso}{NGSO}{Non-Geostationary Satellite Orbits}
\newacronym{mtc}{MTC}{Machine-Type Communications}
\newacronym{nbiot}{NB-IoT}{Narrowband Internet of Things}
\newacronym{etsi}{ETSI}{European Telecommunications Standards Institute}
\newacronym{bsm}{BSM}{Broadband Satellite Multimedia}
\newacronym{haps}{HAPS}{High-Altitude Platform Stations}
\newacronym{embb}{eMBB}{Enhanced Mobile Broadband}
\newacronym{mmtc}{mMTC}{Massive Machine Type Communications}
\newacronym{urllc}{URLLC}{Ultra-Reliable Low-Latency Communications}
\newacronym{ebu}{EBU}{European Broadcasting Union}
\newacronym{dth}{DTH}{Direct-to-Home}
\newacronym{acma}{ACMA}{Australian Communications and Media Authority}
\newacronym{cept}{CEPT}{European Conference of Postal and Telecommunications Administrations}
\newacronym{citel}{CITEL}{Inter-American Telecommunication Commission}
\newacronym{apt}{APT}{Asia-Pacific Telecommunity}
\newacronym{aws}{AWS}{Amazon Web Services}
\newacronym{esa}{ESA}{European Space Agency}
\newacronym{qpsk}{QPSK}{Quadrature Phase Shift Keying}
\newacronym{qam}{QAM}{Quadrature Amplitude Modulation}
\newacronym{tvws}{TVWS}{Television White Spaces}
\newacronym{darpa}{DARPA}{Defense Advanced Research Projects Agency}
\newacronym{iss}{ISS}{International Space Station}
\newacronym{scan}{SCaN}{Space Communications and Navigation}
\newacronym{knn}{KNN}{K-Nearest Neighbours}
\newacronym{pca}{PCA}{Principal Component Analysis}
\newacronym{gpt}{GPT}{Generative Pre-trained Transformer}
\newacronym{gan}{GAN}{Generative Adversarial Network}
\newacronym{fpga}{FPGA}{Field-Programmable Gate Array}
\newacronym{gpp}{GPP}{General-Purpose Processor}
\newacronym{vnf}{VNF}{Virtual Network Function}
\newacronym{mec}{MEC}{Multi Access Edge Computing}
\newacronym{usa}{USA}{United States of America}
\newacronym{uk}{UK}{United Kingdom}
\newacronym{vae}{VAE}{Variational Autoencoder}
\newacronym{los}{LoS}{Line of Sight}
\newacronym{bbm}{BBM}{Break Before Make}
\newacronym{acm}{ACM}{Adaptive Coding and Modulation}
\newacronym{ho}{HO}{Handover}
\newacronym{rb}{RB}{Resource Block}
\newacronym{prach}{PRACH}{Physical Random Access Channel}
\newacronym{isp}{ISP}{Internet Service Provider}
\newacronym{mcs}{MCS}{Modulation and Coding Scheme}
\newacronym{ack}{ACK}{Acknowledgement}

\makeglossaries

\begin{document}
\title{Evaluation of Quantum Machine Learning Models for Network Anomaly Detection}

\author{
\IEEEauthorblockN{Swathi Chandrasekhar and Shiva Raj Pokhrel}
IoT \& SE Research Lab, School of IT, Deakin University, Geelong, Australia\\

\IEEEauthorblockN{Email: swathi.chandrasekhar@deakin.edu.au, shiva.pokhrel@deakin.edu.au}
}
 \maketitle
 

\begin{abstract}
The proliferation of Internet-connected devices has intensified the risk of large-scale network intrusions, necessitating more advanced detection methodologies for the post quantum era. This work presents, to the best of our knowledge, a comprehensive investigation of quantum machine learning (QML) for anomaly detection on the BoT-IoT dataset, a widely adopted benchmark for IoT security. Leveraging feature-engineered representations, we implement and evaluate four representative quantum classifiers: Quantum Support Vector Classification (QSVC), Quantum Neural Networks (QNN), and Variational Quantum Classifiers (VQC). 

\end{abstract}

\begin{IEEEkeywords}
Quantum Machine Learning (QML), Anomaly Detection.
\end{IEEEkeywords}

\section{Introduction}

The proliferation of digital technologies and the Internet of Things (IoT) has generated vast, high-dimensional datasets in domains such as consumer electronics, finance, and critical infrastructure. As these systems become more interconnected, the challenge of detecting anomalies—ranging from device failures to financial fraud—has grown increasingly urgent. While classical machine learning (ML) has achieved notable success in anomaly detection, it often struggles with scalability, robustness in high-dimensional spaces, and generalization to novel threats~\cite{bhowmik2024quantum}. Quantum computing, leveraging superposition and entanglement, offers a fundamentally new computational paradigm that can surpass classical limits~\cite{liu2018quantum,kyriienko2022unsupervised,Chandrasekhar2025}. Within this paradigm, quantum machine learning (QML)~\cite{corli2025quantum, kumari2025modelingwavelettransformedquantum} has shown promise for complex, high-dimensional, and noisy data~\cite{bhowmik2024quantum,kyriienko2022unsupervised,hdaib2024quantum}, with emerging evidence suggesting its potential to efficiently identify rare and subtle anomalies.

Hybrid quantum-classical models have demonstrated early success in IoT and consumer electronics by capturing intricate, non-linear correlations~\cite{bhowmik2024quantum}. In finance, quantum kernel methods have outperformed classical RBF kernels as feature dimensionality increases, especially in unsupervised settings~\cite{kyriienko2022unsupervised}. Variational quantum circuits and quantum support vector machines further mitigate overfitting and enhance generalization in heterogeneous and noisy data~\cite{kyriienko2022unsupervised}.

Despite these advances, practical deployment of QML-based anomaly detection faces challenges, including the computational cost of large quantum kernel matrices and hardware limitations such as qubit count and gate fidelity~\cite{hdaib2024quantum,singh2025modeling,preskill2018quantum}. Ongoing research into efficient kernel evaluation, sparse matrix techniques, and hardware-specific optimizations is gradually improving feasibility~\cite{kyriienko2022unsupervised}.

In this work, we systematically investigate the potential and limitations of QML for anomaly detection in real-world domains. By benchmarking quantum and classical methods on representative datasets, we delineate the conditions for quantum advantage and highlight key algorithmic and hardware bottlenecks. Our discoveries offer new insights into the expressivity, robustness, and scalability of quantum-enhanced anomaly detection, guiding future progress toward practical quantum advantage in security-critical, data-intensive environments.

\subsection{Key Contributions}

\begin{itemize}
    \item We propose a unified QML framework for anomaly detection, instantiated and evaluated on the BoT-IoT benchmark dataset\cite{peterson2021review}, providing the first systematic study of quantum enhanced methods in this security critical domain.
    \item We rigorously assess the influence of dimensionality reduction techniques including Behavioral and Statistical Analysis (BSA), principal component analysis (PCA), and t-distributed stochastic neighbor embedding (t-SNE) on the performance and robustness of quantum classifiers.
    \item We implement and benchmark four state-of-the-art QML models Quantum Support Vector Classifier (QSVC), Variational Quantum Classifier (VQC), Quantum Neural Network (QNN) for anomaly detection in IoT network traffic.
    \item We provide a comparative performance evaluation of these quantum approaches against classical baselines, identifying the conditions under which QML models yield advantages, and analyze their scalability, generalization, and deployment feasibility for real-world IoT security applications.
\end{itemize}

\section{Methodology}





We discover a novel framework for implementing and evaluating QML algorithms for anomaly detection in IoT network traffic, using the BoT-IoT dataset. The framework comprises two main stages: (i) dimensionality reduction and feature selection, and (ii) quantum-based classification.

\textbf{Feature Selection and Dimensionality Reduction:}  
The dataset is first preprocessed to identify the most relevant features. Feature importance is quantified using behavioral and statistical metrics, including point-biserial correlation, chi-square tests, and ANOVA F-tests, with aggregated scores used to rank and select top features. For linear dimensionality reduction, we apply Principal Component Analysis (PCA)~\cite{singh2025modelinggenomic}, which projects data onto a lower-dimensional subspace by solving the eigenvalue problem: $\mathbf{C} \mathbf{v}_i = \lambda_i \mathbf{v}_i$, where $\mathbf{C}$ is the covariance matrix, $\mathbf{v}_i$ are eigenvectors, and $\lambda_i$ are eigenvalues.

To capture non-linear structure, we employ t-distributed Stochastic Neighbor Embedding (t-SNE), which preserves local similarities by minimizing the Kullback--Leibler divergence between high- and low-dimensional probability distributions: $\mathrm{KL}(P \| Q) = \sum_{i \neq j} p_{ij} \log \frac{p_{ij}}{q_{ij}}$, where $p_{ij}$ and $q_{ij}$ denote pairwise similarities in the original and embedded spaces, respectively.

\textbf{Quantum Classification:}  
The reduced feature sets are then used as input to several quantum classifiers. Each model exploits different quantum mechanisms to enhance discrimination power in high-dimensional and noisy regimes.

\subsection{Quantum Support Vector Classifier (QSVC)}  
QSVC extends classical support vector machines by employing quantum kernels that map data into a high-dimensional Hilbert space~\cite{corli2025quantum, singh2025modeling,chandrasekhar2025modelingquantumautoencodertrainable}. The kernel function between two input vectors $x$ and $x'$ is expressed as $K(x, x'; \theta) = \left| \langle \psi(x, \theta) \mid \psi(x', \theta) \rangle \right|^2$, where $\ket{\psi(x, \theta)}$ is the quantum state generated by a parameterized feature map. This formulation enables QSVC to capture complex correlations inaccessible to classical kernels, particularly in high-dimensional IoT traffic data.

\begin{algorithm}
\caption{Quantum Support Vector Classifier (QSVC)}
\label{alg:qsvc}
\begin{algorithmic}[1]
\Procedure{QSVC\_Train}{$\{(\vec{x}_i, y_i)\}_{i=1}^N$, feature\_map, optimizer}
    \State Apply dimensionality reduction to $\vec{x}_i$
    \State Normalize features
    \State Initialize parameters $\theta$
    \For{each pair $(\vec{x}_i, \vec{x}_j)$}
        \State Prepare quantum states $|\psi(\vec{x}_i, \theta)\rangle$, $|\psi(\vec{x}_j, \theta)\rangle$
    \EndFor
    \State Optimize $\theta$ to minimize SVC loss
    \State Train QSVC with optimized kernel $K$ and labels $y_i$
    \State \textbf{return} trained QSVC model
\EndProcedure
\end{algorithmic}
\end{algorithm}

\subsection{Variational Quantum Classifier (VQC)}
VQC is a hybrid quantum-classical algorithm that uses a parameterized quantum circuit and a classical optimizer to learn a decision boundary \cite{Chandrasekhar2025,chandrasekhar2025modelingquantumautoencodertrainable}. The model minimizes:
\[
\min_{\theta} \; \mathcal{L}(\theta) = \sum_{i=1}^N \ell\left(f_{\theta}(x_i), y_i\right)
\]
where $f_{\theta}(x_i)$ is the predicted probability and $\ell$ is a loss function.

\begin{algorithm}
\caption{Variational Quantum Classifier (VQC)}
\label{alg:vqc}
\begin{algorithmic}[1]
\Procedure{VQC\_Train}{$\{(\vec{x}_i, y_i)\}_{i=1}^N$, feature\_map, ansatz, optimizer}
    \State Apply dimensionality reduction to $\vec{x}_i$ 
    \State Normalize features and Initialize parameters $\theta$
    \For{each training iteration}
        \For{each $\vec{x}_i$}
            \State Prepare quantum state with feature\_map,ansatz
            \State Compute predicted probabilities $f_{\theta}(\vec{x}_i)$
        \EndFor
        \State Compute loss $\mathcal{L}(\theta)$
    \EndFor
    \State \textbf{return} trained VQC model
\EndProcedure
\end{algorithmic}
\end{algorithm}

\subsection{Quantum Neural Network (QNN)}
QNNs are parameterized quantum circuits that approximate complex functions, analogous to classical neural networks \cite{liu2018quantum, Chandrasekhar2025, singh2025modeling,chandrasekhar2025modelingquantumautoencodertrainable}. The QNN output is the expectation value:
\[
f_{\theta}(x) = \langle 0 | U^\dagger(x, \theta) \hat{O} U(x, \theta) | 0 \rangle
\]
where $U(x, \theta)$ is the parameterized circuit and $\hat{O}$ is the measurement observable.

\begin{algorithm}
\caption{Quantum Neural Network (QNN)}
\label{alg:qnn}
\begin{algorithmic}[1]
\Procedure{QNN\_Train}{$\{(\vec{x}_i, y_i)\}_{i=1}^N$, feature\_map, ansatz, optimizer}
    \State Apply dimensionality reduction to $\vec{x}_i$
    \State Normalize features and Initialize parameters $\theta$
    \For{each training iteration}
        \For{each $\vec{x}_i$}
            \State Prepare quantum state with feature\_map,ansatz
            \State Compute output $f_{\theta}(\vec{x}_i)$
        \EndFor
        \State Compute loss $\mathcal{L}(\theta)$
    \EndFor
    \State \textbf{return} trained QNN model
\EndProcedure
\end{algorithmic}
\end{algorithm}

To instantiate our framework, we use the BoT-IoT dataset preprocessed following~\cite{peterson2021review}. From the original traffic traces, we retain flow-level statistical and behavioral features (e.g., packet counts, byte rates, inter-arrival times, and flag-based indicators), yielding 20 features after BSA-based ranking. The resulting dataset is randomly partitioned into 70\% training, 15\% validation, and 15\% test sets, stratified by attack/benign labels to preserve class proportions. All features are z-score normalized prior to dimensionality reduction (PCA or t-SNE) and subsequent quantum encoding. Quantum circuits are simulated using Qiskit Aer on a Mac-OS, ensuring that performance comparisons reflect realistic near-term execution costs for QML-based anomaly detection.

\section{Results and Experimentation}

\begin{table*}[ht]
\centering
\caption{Performance of Classical ML and QML Algorithms for Anomaly Detection on the BoT-IoT Dataset Using Different Dimensionality Reduction Techniques.}
\label{tab:qml_results}
\begin{threeparttable}

\begin{tabular}{|c|c|c|c|c|c|c|}
\hline
\textbf{Algorithm} & \textbf{Accuracy} & \textbf{Precision} & \textbf{F1-score} & \textbf{Recall} & \textbf{Time (s)}  &  \textbf{Dimensionality reduction Technique}  \\
\hline
SVC (classical ML) & 0.87 & 0.71 & 0.83 & 0.88 & 268 & BSA \\
\hline
QSVC & 0.85 & 0.72 & 0.85 & 0.78 & 736 & BSA \\
\hline
QNN & 0.85 & 0.85 & 0.72 & 0.92 & 164 & BSA \\
\hline
VQC & 0.95 & 0.96 & 0.95 & 0.95 & 266 & BSA \\
\hline
SVC (classical ML) & 0.97 & 0.91 & 0.93 & 0.88 & 141 & PCA \\
\hline
QSVC & 0.94 & 0.97 & 0.95 & 0.94 & 2152 & PCA \\
\hline
QNN & 0.74 & 0.71 & 0.83 & 0.79 & 472 & PCA \\
\hline
VQC & 0.52 & 0.64 & 0.49 & 0.58 & 2230 & PCA \\
\hline
SVC (classical ML) & 0.77 & 0.81 & 0.76 & 0.78 & 285 & t-SNE \\
\hline
QSVC & 0.89 & 0.80 & 0.91 & 0.94 & 872 & t-SNE \\
\hline
QNN & 0.62 & 0.67 & 0.62 & 0.73 & 130 & t-SNE \\
\hline
VQC & 0.74 & 0.73 & 0.74 & 0.85 & 132 & t-SNE \\
\hline
\end{tabular}
\begin{tablenotes}
\small
\item \textbf{Abbreviations:} QSVC = Quantum Support Vector Classifier; QNN = Quantum Neural Network; VQC = Variational Quantum Classifier; BSA = Behavioral and Statistical Analysis; PCA = Principal Component Analysis; t-SNE = t-distributed Stochastic Neighbor Embedding.
\end{tablenotes}
\end{threeparttable}
\end{table*}

Table~\ref{tab:qml_results} presents the performance of three QML algorithms (QSVC, QNN, and VQC) and classical SVC on the BoT-IoT dataset with three dimensionality reduction techniques: BSA, PCA, and t-SNE. Metrics include accuracy, precision, recall, F1-score, and computational time. VQC with PCA achieves the highest accuracy and F1-score, while QSVC offers strong precision with lower computational cost. Training loss visualizations (Figures~\ref{fig:bs_loss}–\ref{fig:violin_loss}) indicate that VQC converges faster and more stably across dimensionality reduction methods, suggesting its promise for scalable IoT anomaly detection.

Classical SVC generally matches or outperforms QML models in both accuracy and efficiency. Nonetheless, QML algorithms already show competitive results, and as quantum hardware and algorithms mature, they are expected to surpass classical methods, particularly for complex, high-dimensional data.

\subsection{Behavioral and Statistical Analysis (BSA)}
\begin{figure}[h]
  \centering
  \includegraphics[width=0.8\columnwidth]{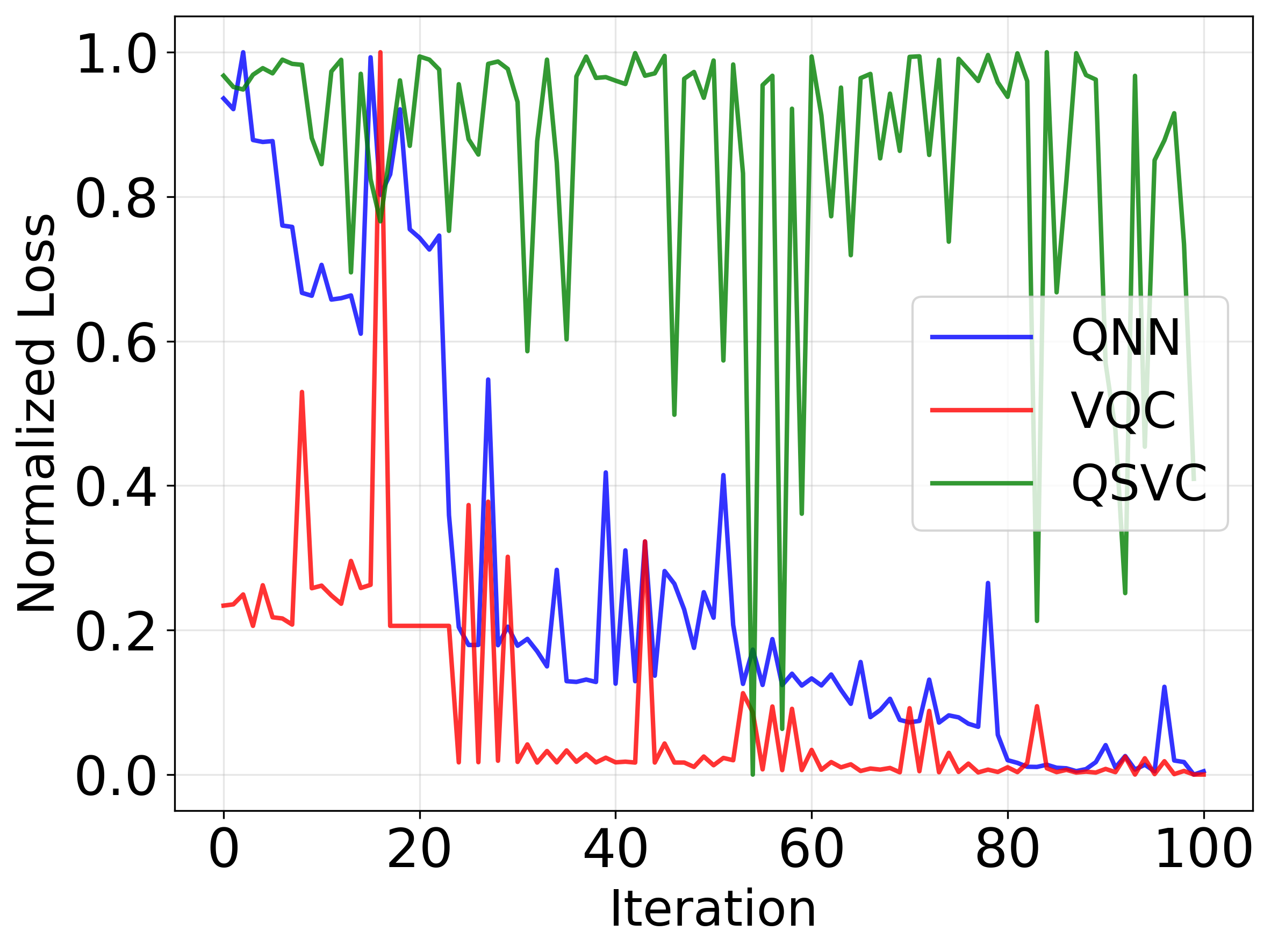}
  \caption{Normalized training loss across iterations for QNN, VQC, and QSVC using BSA.}
  \label{fig:bs_loss}
\end{figure}

Figure~\ref{fig:bs_loss} illustrates the convergence trajectories of QNN, VQC, and QSVC using BSA features. VQC exhibits rapid loss minimization, approaching near-zero within 30 iterations, consistent with its top performance (accuracy = 0.95, F1 = 0.95). QNN converges more gradually over ~50 iterations, reflecting its robust recall (0.92), while QSVC displays pronounced oscillations between 0.6–1.0, indicating unstable training despite moderate accuracy (0.85). These results suggest that BSA-derived representations are more effectively leveraged by parameterized quantum circuits (VQC, QNN) than by kernel-based methods.

\subsection{Principal Component Analysis (PCA)}
\begin{figure}[h]
  \centering
  \includegraphics[width=0.8\columnwidth]{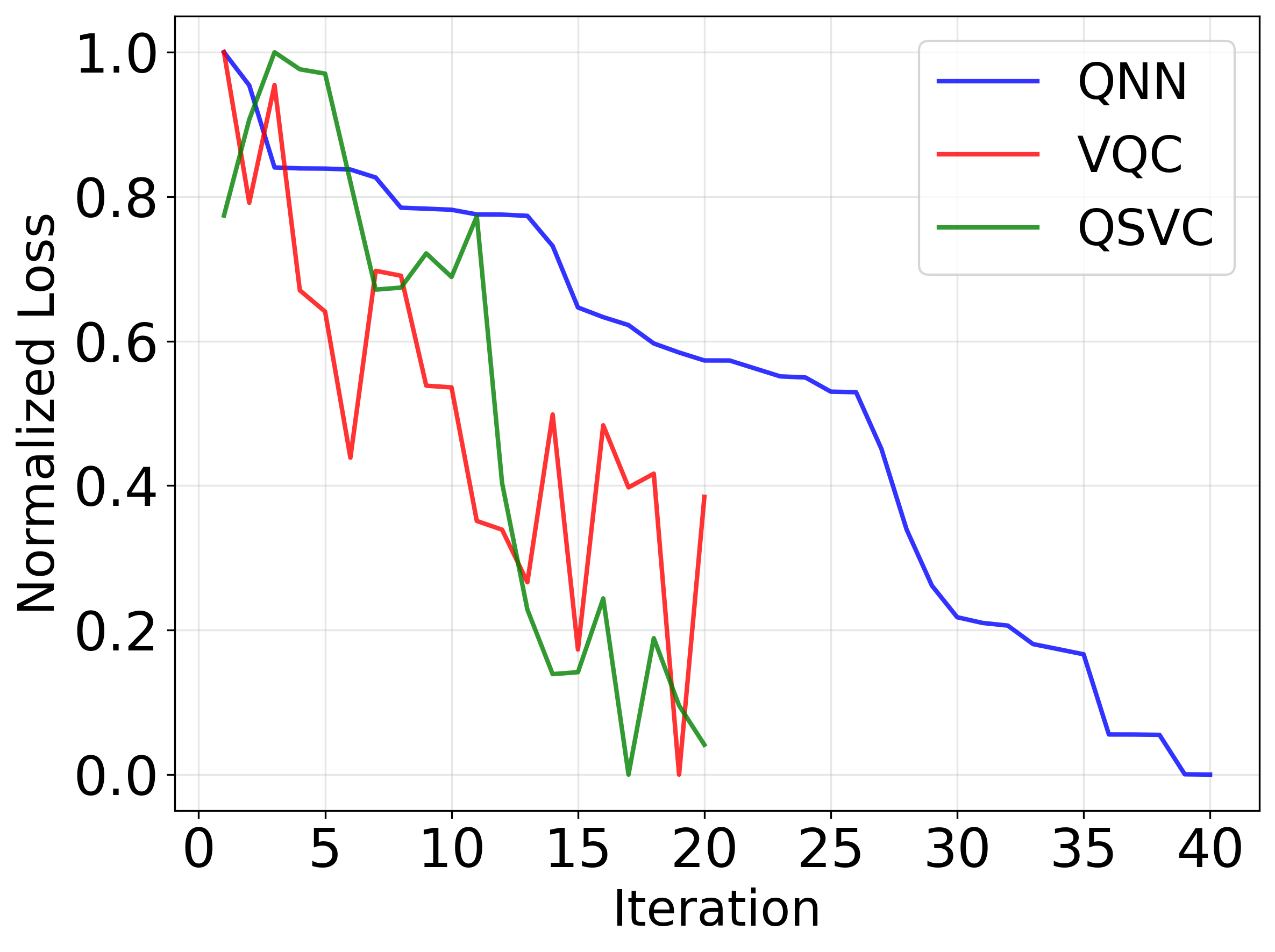}
  \caption{Normalized training loss across iterations for QNN, VQC, and QSVC using PCA.}
  \label{fig:pca_loss}
\end{figure}

Figure~\ref{fig:pca_loss} shows the convergence behavior of QSVC, QNN, and VQC on PCA-reduced features. QSVC exhibits smooth and consistent convergence, achieving the highest accuracy (0.94) and F1-score (0.95), highlighting its suitability for linearly compressed representations. QNN converges slowly, suggesting limited expressivity on PCA-transformed features, while VQC experiences sharp loss spikes and unstable training, resulting in poor accuracy (0.52). Notably, both QSVC and VQC incur high computational costs (>2000\,s), indicating that while PCA can enhance QSVC performance, it also substantially increases runtime.

\subsection{t-distributed Stochastic Neighbor Embedding (t-SNE)}
\begin{figure}[h]
  \centering
  \includegraphics[width=0.8\columnwidth]{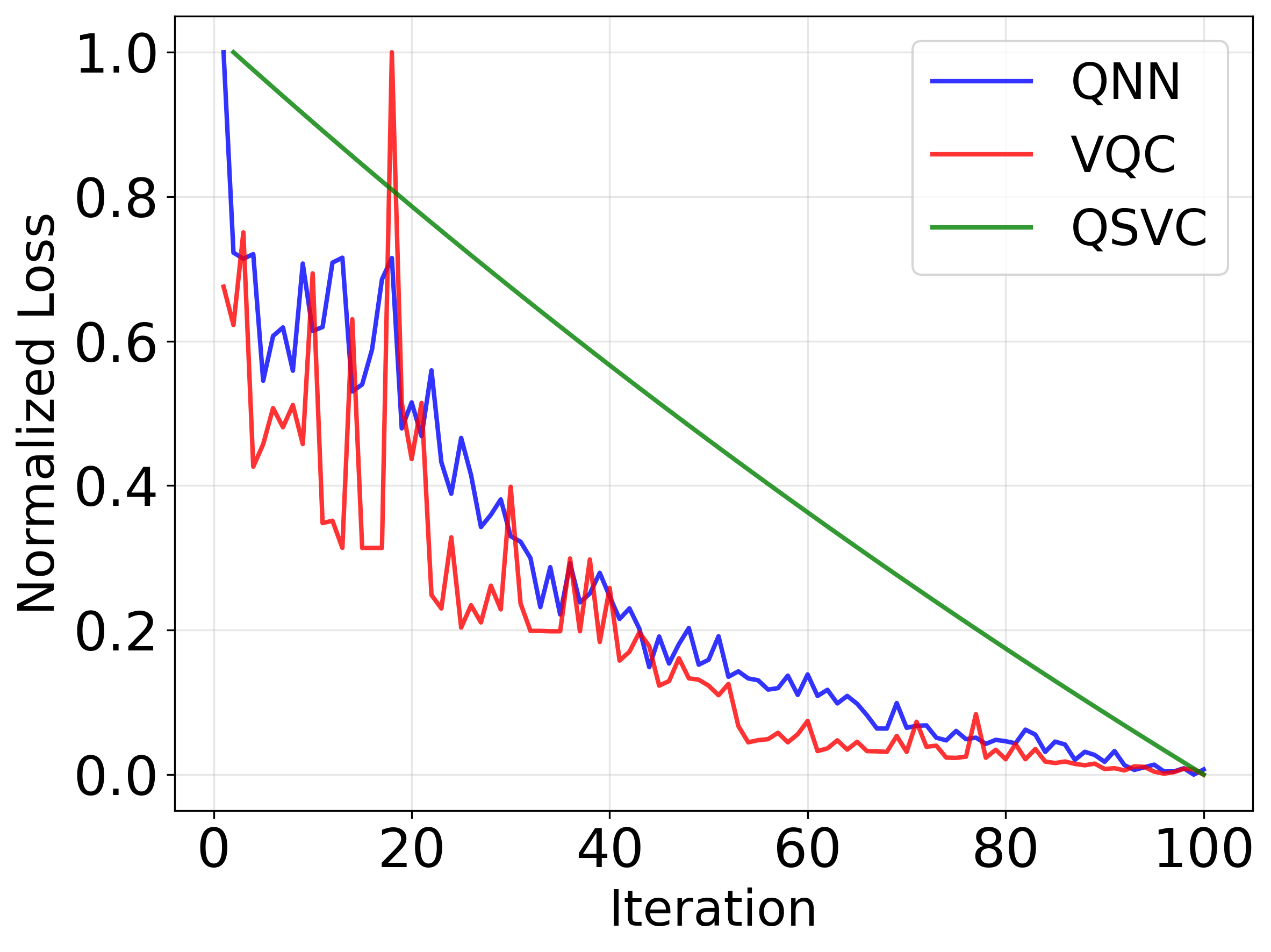}
  \caption{Normalized training loss across iterations for QNN, VQC, and QSVC using t-SNE.}
  \label{fig:tsne_loss}
\end{figure}

Figure~\ref{fig:tsne_loss} illustrates the training dynamics of QSVC, QNN, and VQC on t-SNE features. QSVC demonstrates nearly linear, monotonic convergence from loss = 1.0 to 0.0, consistent with its strong recall (0.94) and robust F1-score (0.91). QNN and VQC exhibit oscillatory convergence, stabilizing after ~100 iterations, with VQC ultimately achieving lower final loss and higher accuracy (0.74 vs. 0.62). Notably, t-SNE significantly improves computational efficiency, reducing runtime for QNN (130\,s) and VQC (132\,s) compared to PCA-based training, highlighting its effectiveness for fast, high-dimensional anomaly detection.

\subsection{Loss Distribution Across Models and Dimensionality reduction Techniques}
\begin{figure}[h]
  \centering
  \includegraphics[width=0.8\columnwidth]{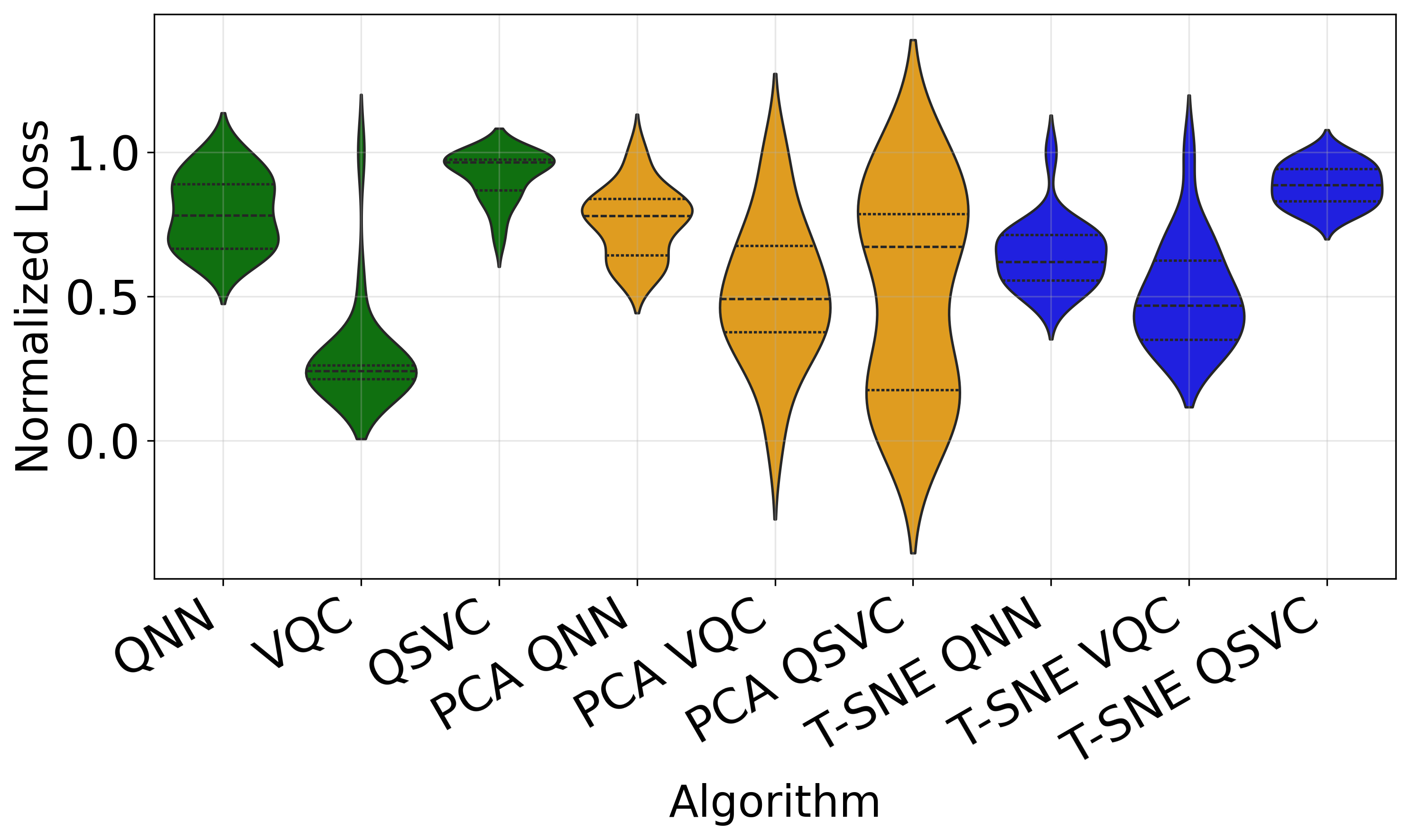}
  \caption{Distribution of normalized losses for QNN, VQC, and QSVC under different Dimensionality reduction techniques (BSA, PCA, t-SNE).}
  \label{fig:violin_loss}
\end{figure}

Figure~\ref{fig:violin_loss} presents violin plots of normalized loss distributions for all algorithm–Dimensionality reduction combinations. VQC with BSA exhibits the narrowest, lowest-centered distribution, confirming both stable training and superior predictive performance. QSVC on PCA, despite high accuracy, shows the widest spread, reflecting substantial variability across iterations. t-SNE yields more balanced distributions for all algorithms, with QSVC maintaining consistently lower spread than QNN and VQC. Overall, these results indicate that BSA optimally stabilizes VQC, PCA favors QSVC, and t-SNE offers a compromise between computational efficiency and moderate reliability.

\subsection{Comparative Analysis}
Integrating the quantitative outcomes (Table~\ref{tab:qml_results}) with convergence and distribution analyses (Figures~\ref{fig:bs_loss}–\ref{fig:violin_loss}), three conclusions emerge:
\begin{itemize}
    \item \textbf{VQC + BSA} achieves the most favorable trade-off between stability, accuracy, and computational time.
    \item \textbf{QSVC + PCA} delivers the highest accuracy but incurs significant computational costs and variance in loss.
    \item \textbf{QSVC + t-SNE} provides strong recall with smooth convergence and efficient runtimes, making it attractive for resource-constrained deployments.
\end{itemize}

\section{Conclusion}
This paper presented a comparative evaluation of QML algorithms for anomaly detection on the BoT-IoT dataset using different dimensionality reduction techniques. Our results show that VQC combined with BSA achieved the best overall trade-off between accuracy, stability, and runtime, while QSVC with PCA delivered the highest accuracy at the cost of significant computational time. QSVC with t-SNE provided efficient convergence with strong recall, making it suitable for resource-limited IoT environments. In practice, t-SNE driven feature compression reduces circuit depth and evaluation time, which is particularly beneficial when QML inference is offloaded from cloud backends to edge gateways or lightweight security appliances, thereby aligning our design with the stringent latency and resource constraints typical of IoT deployments. QNN demonstrated robustness in recall but exhibited greater variability across reductions.

\vspace{-1mm}


\bibliographystyle{IEEEtran} 
\bibliography{mybib}

\end{document}